\documentclass{aa} 
\usepackage{graphicx}
\usepackage{txfonts}
\usepackage{natbib}

\newcommand{\be}{\begin{equation}}
\newcommand{\ee}{\end{equation}}

\begin{document} 
 
\title
{The importance of the merging activity for the kinetic polarization
of the Sunyaev-Zel'dovich signal from galaxy clusters}
\titlerunning{Merging activity and the kinetic SZ polarization from galaxy clusters}

\author{Matteo Maturi\inst{1}
  \and Lauro Moscardini\inst{2,3}
  \and Pasquale Mazzotta\inst{4,5}
  \and Klaus Dolag\inst{6}
 \and Giuseppe Tormen\inst{7}
}

\institute{Zentrum f\"ur Astronomie, ITA, Universit\"at 
   Heidelberg, Albert-\"Uberle-Str.~2, D-69120 Heidelberg, Germany
   \and
   Dipartimento di Astronomia, Universit\`a di Bologna, via 
   Ranzani 1, I-40127 Bologna, Italy
   \and
   INFN - National Institute for Nuclear Physics, Sezione di Bologna, 
   viale Berti Pichat 6/2,  I-40127 Bologna, Italy
   \and
   Dipartimento di Fisica, Universit\`a di Roma Tor
   Vergata, via della Ricerca Scientifica 1, I-00133 Roma, Italy
   \and
   Harvard-Smithsonian Center for Astrophysics, 60 Garden Street, Cambridge, MA  02138, USA
   \and
   Max-Planck Institute fuer Astrophyik, Karl-Schwarzschild Strasse 1, 
   D-85748 Garching, Germany 
   \and
    Dipartimento di Astronomia, Universit\`a di Padova, vicolo dell'Osservatorio 2, 
    I-35122 Padova, Italy
}

\date{\emph{Astronomy \& Astrophysics, to be submitted}} 

\abstract
{The polarization sensitivity of the upcoming millimetric
observatories will open new possibilities for studying the properties
of galaxy clusters and for using them as powerful cosmological probes.
For this reason it is necessary to investigate in detail the
characteristics of the polarization signals produced by their highly
ionized intra-cluster medium (ICM).  This work is focussed on the
polarization effect induced by the ICM bulk motions, the so-called
kpSZ signal, which has an amplitude proportional to the optical depth
and to the square of the tangential velocity.}
{In particular we study how this polarization signal is affected by
the internal dynamics of galaxy clusters and what is its dependence on
the physical modelling adopted to describe the baryonic component.}
{This is done by producing realistic kpSZ maps starting from the
outputs of two different sets of high-resolution hydrodynamical N-body
simulations. The first set (17 objects) follows only non-radiative
hydrodynamics, while for each of 9 objects of the second set we
implement four different kinds of physical processes.}
{Our results shows that the kpSZ signal turns out to be a very
sensitive probe of the dynamical status of galaxy clusters. We find
that major merger events can amplify the signal up to one order of
magnitude with respect to relaxed clusters, reaching amplitude up to
about 100 $nK$.  This result implies that the internal ICM dynamics
must be taken into account when evaluating this signal because
simplicistic models, based on spherical rigid bodies, may provide
wrong estimates.  In particular, the selection of enough relaxed
clusters seems to be fundamental to obtain a robust measurement of the
intrinsic quadrupole of the cosmic microwave background through
polarization.  Finally we find that the dependence on the physical
modelling of the baryonic component is relevant only in the very inner
regions of clusters.}
{}


\maketitle

\section{Introduction}

The last generation of X-ray satellites like Chandra and XMM-Newton
have shown a complex scenario for the intracluster medium (ICM). In
particular the presence of cold fronts has been detected in many
galaxy clusters \citep[see, e.g.,][]{allen2001}. The existence of this
phenomenon suggests that also relaxed clusters display a picture in
contrast with the previously hypothesized hydrostatic state for the
ICM, showing minor merger events and evident gas motions.

In this paper we focus our attention on a different signal which could
give strong constraints on the merging activity of galaxy clusters:
the kinematic Sunyaev-Zel'dovich (SZ) polarization (hereafter $kpSZ$)
\citep{sunyaev1980b}. It takes origin by single scattering of moving
electrons, and its dependence on the square of their tangential
velocity makes it a suitable probe for investigating the dynamic
properties of clusters during the non-linear process of their
formation.

As we will discuss in more detail in Sect.~\ref{sec:polarizations},
galaxy clusters produce other linear polarization contributions to the
cosmic microwave background (CMB) radiation through different
processes \citep[see, e.g.,][]{sazonov1999}. Basically all components
of the quadrupole moment present in the incoming radiation induce a
linear polarization: in fact there is the polarization induced by the
intrinsic quadrupole of the CMB radiation ($P_Q$), by the quadrupole
component of kinetic and thermal SZ effects ($P_{kp^2SZ}$ and
$P_{tp^2SZ}$, respectively) and by the gravitational lensing effect
($P_l$) of clusters them-self. These polarization contributions, with
the exception of $P_{kp^2SZ}$, which is one order of magnitude
smaller, have amplitudes similar to the kpSZ effect
\citep{lavaux2004}. Thus, none of them can be neglected. In any case,
a component separation is possible, thanks to their different
frequency dependences.  It is important to notice that each of these
signals carries different and important information on galaxy clusters
and cosmology. For example, $P_Q$ could be used to estimate the
intrinsic quadrupole at different redshift and thus to investigate the
dark energy evolution; $P_{tp^2SZ}$ can be used to constrain the gas
profile thanks to its dependence on the square of the gas density;
$P_{kpSZ}$, which we are going to discuss in more detail in this
paper, is a tool to study the ICM dynamics and thus the history of the
merging activity.

Using a toy model, \cite{diego2003} showed that the kpSZ effect
strongly depends on sub-clump motions, and that major merger events
could make this signal equal or dominant on the other ones (depending
on the assumed frequency). A different study based on a numerical
simulation has been carried out by \cite{lavaux2004}, who focussed
their attention on the large scale structure of the kpSZ signal. Their
results show that the environmental collapse into the potential of the
forming cluster originates a characteristic kpSZ pattern with circular
symmetry. They suggested to use this peculiar pattern to detect
forming clusters.  In addition they confirmed the analytic results of
\cite{sazonov1999}, but they did not investigate the dependences of
this signal on the merging activity of the clusters and they did not
perform any extended statistical study. \cite{shimon2006}, using a
cluster simulated by the {\it ENZO} code, investigate the polarization
levels and patterns produced by the different components originated by
the kinematic quadrupole moments induced by moving scattering
electrons: their conclusion is that the signal is high enough
($\approx 1\mu$K) to become detectable in upcoming CMB experiments.

Here we extend these previous works by analyzing two different sets of
high-resolution hydrodynamical N-body numerical simulations, performed
by assuming different modeling for the ICM physics, to investigate in
detail the statistical properties of the kpSZ signal. We demonstrate
that it can vary up to one order of magnitude according to the
internal cluster dynamics.

The plan of this paper is as follows. In Sect.~\ref{sec:polarizations}
we introduce the Stokes parameters and the primary polarization of the
CMB, then we present the main polarization contributions induced by
galaxy clusters, discussing in more detail the kpSZ polarization.
Sect.~\ref{sec:pol_sim} is devoted to an extended description of the
two sets of numerical hydrodynamical simulations used for our
analysis.  Sect.~\ref{sec:examples} introduces two different examples
of galaxy clusters, a relaxed cluster participating to a major merger,
and a dynamically more active object: the corresponding signals are
compared to investigate the influence of the in-falling haloes on the
kpSZ signal.  The statistical analysis of the full samples of
simulated objects is presented in Sect.~\ref{sec:statistic}, where we
consider separately the results of non-radiative simulations and the
ones obtained including different ICM modeling.  Finally, our main
conclusions are summarized in Sect.~\ref{sec:pol_conclusion}.

\section{Linear polarization by scattering with free electrons}
\label{sec:polarizations}

Scattering by free electrons induces a linear polarization component
only if the incoming radiation has a quadrupole anisotropy. In the
case of the CMB photons, we can distinguish different polarization
components according to the origin of the quadrupole moment of the
incoming radiation and to the location of the free electrons. In this
section we briefly review these processes after having introduced the
main quantities related to polarization.

\subsection{Parametrizing polarization}

Radiation can be parametrized by using the four Stokes parameters: $I$
represents the total intensity of the light, $Q$ and $U$ quantify the
degree of linear polarization, while $V$ gives the degree of circular
polarization. Since the Thompson scattering does not introduce any
circular polarization, we can set $V=0$ here. Notice that polarization
is a pseudo-vector (i.e. it has not a verse), whose direction is given
by
\be
  \chi = \frac{1}{2} \tan^{-1}\left(\frac{U}{Q}\right) \,.
\ee
The degree of polarization $p$ can be expressed as
\be
  p=\frac{\sqrt{Q^2+U^2}}{I} \,.
\ee

The Stokes parameters are additive along the line-of-sight, and thus
the total effect can be computed as the integral of local effects:
$Q=\int dQ$ and $U=\int dU$.

Throughout this paper, we compute the polarization effect assuming the
incoming CMB radiation to be unaffected by the SZ effect,
i.e. $I\simeq I_{\rm CMB}$: this is justified by the assumption of low
optical depth.

\subsection{Primary CMB polarization}

The CMB has a primordial polarization component originated at the last
scattering surface. At that epoch, the universe was fully ionized and
the quadrupole moment was related to the primary anisotropies.  This
signal is expected to peak inside the cosmological horizon scale at
that time, because it is a causal process requiring a coherence into
the electron motion. This polarization signal has been recently
measured by the WMAP satellite, which also detected a large-scale
component $l (l+1) C_{l=\langle 2-6\rangle}^{\rm
EE}/2\pi=0.086\pm0.029 \mu {\rm K}^2$, interpreted as a signature of a
global reionization occurred at redshift $z=10.9^{+2.7}_{-2.3}$
\citep{page2006}.

\subsection{Secondary CMB polarization by galaxy clusters}

Free electrons are also present into the ICM of galaxy clusters. In
this case the polarization components depend on the nature of the
quadrupole moment of the incoming radiation, as we are going to
discuss in the following subsections.

\subsubsection{Polarization by single scattering of moving electrons: $pkSZ$}

Our work will be focused on the kpSZ effect which is discussed here in
more detail. This effect arises from single scattering of the CMB
photons by the ICM electrons. A moving electron sees the CMB radiation
Doppler-shifted because of its peculiar motion and thus perceives a
radiation intensity given by

\be
  I(\nu,\mu) = \frac{C x^3}{e^{x\gamma(1+\beta\mu)}-1}\,,
\ee
where $x\equiv h\nu/(kT)$ is the adimensional frequency, $\beta\equiv
v_e/c$ gives the electron velocity in units of the speed of light,
$\gamma\equiv(1-\beta^2)^{-1/2}$ is the Lorentz factor, $\mu$ is the
cosine of the angle between the direction of the electron velocity and
the one of the incident CMB photon. The decomposition of this function
into Legendre polynomials leads to
\be\label{eqn:I_qudrupole}
  I(\nu,\mu) = \frac{C x^3}{e^x-1}
                     \left[
		       I_0 +
		       I_1 \mu +
		       I_2 \left( \mu^2 -\frac{1}{3}\right)
		       \frac{e^{x}(e^{x}+1)}{2(e^{x}-1)^2}x^2 \beta^2 +
		       ...
		     \right] \,.
\ee

The previous equation contains the quadrupole component necessary to
produce a linear polarization orthogonal to the electron velocity
direction \citep{sunyaev1980,sazonov1999}.

The resulting local contributions to the $Q$ and $U$ Stokes parameters
are
\be\label{eqn:dQ_dU}
\begin{array}{c}
  \displaystyle
  dQ =-0.1 \sigma_T n_e f(x) \beta_t^2 \cos(2\chi) dl \\
  \displaystyle
  dU =-0.1 \sigma_T n_e f(x) \beta_t^2 \sin(2\chi) dl
\end{array}
\,,
\ee 
respectively. In the previous equation, $\sigma_T$ represents the
Thompson cross-section, $n_e$ is the electron density, $\beta_t$ is
the component of $\vec{\beta}$ on the sky plane (tangential velocity),
$\chi$ defines the position angle of the tangential velocity, and
\be
  f(x) = \frac{e^{x}(e^{x}+1)}{2(e^{x}-1)^2}x^2 
\ee
is the non-relativistic approximation of the frequency dependence of
the polarization signal, as results of the quadrupole component of
Eq.(\ref{eqn:I_qudrupole}). The Stokes parameters (see
Eq.\ref{eqn:dQ_dU}) are normalized by the incoming mean intensity of
the CMB radiation and are referred to the observer rest frame. They
can be converted into a brightness temperature by dividing the
previous quantities by
\be
  g(x)=\frac{d \,\ln I_\nu}{d \, \ln T} = \frac{x e^x}{e^x-1}\,.
\ee
Eq.~(\ref{eqn:dQ_dU}) shows the kpSZ effect dependence on the square
of the tangential velocity $\beta_t^2$ and on its direction
$\chi$. Therefore this effect is very sensitive to the ICM dynamics
and its amplitude can be used to investigate the merging activity.

\subsubsection{Polarization by single scattering of the intrinsic CMB
               quadrupole: $P_{Q}$}

The CMB radiation has an intrinsic quadrupole moment given by the
Sachs-Wolfe effect and by the late Integrated Sachs-Wolfe effect,
which is mostly related to the presence of a dark energy component.
This polarization component depends on the optical depth $\tau$ and on
the amplitude of the quadrupole $Q(z)$ at the cluster redshift, where
the scattering process occurs: $P_{Q} \propto \tau Q(z)$.

Because of that, if we have an estimate of $\tau$, for example thanks
to SZ measurements, we can evaluate the intrinsic quadrupole of the
CMB at different redshifts.  For a flat $\Lambda$CDM model with
$\Omega_\Lambda = 0.7$, the maximum signal is of the order of $4.9
\tau \ \mu\mbox{K}$. Assuming a typical value of $\tau\sim 0.03$, this leads
to a maximum signal of $\sim 0.14 \ \mu\mbox{K}$, as found
analytically by \cite{sazonov1999}.  This result has been confirmed by
the numerical simulations performed by \cite{amblard2005}. A more
detailed study of the primordial CMB quadrupole-induced polarization
based on the combination of an analytic method with hydrodynamic
simulations has been carried out by \cite{liu2005}, who also
investigated the contributions to the signal of different gas phases.
Because of the dependence on $Q$, there are four orthogonal regions
into the sky where $P_{Q}$ vanishes or at least is very low. These
regions have been located by the WMAP satellite: one of them is at
galactic coordinate $(l,b)\approx(-80^\circ,60^\circ)$, close to the
Virgo cluster \citep{tegmark2003}.

This observable is one of the most sensitive probe of the dark energy:
in fact the growth of the gravitational potential, and thus of $Q$, is
directly proportional to its equation of state. On the contrary all
other related quantities depend at least on an integral of this
quantity.

\subsubsection{Polarization by second scattering of the tSZ and kSZ
               quadrupoles: $P_{p^2tSZ}$, $P_{p^2kSZ}$}

A first Thompson scattering of the CMB radiation with the ICM
electrons introduces two secondary anisotropies, called kinetic and
thermal Sunyaev Zel'dovich effects. They have a quadrupole component
and thus a second scattering would originate a linear polarization
related to each of them. We call $P_{p^2tSZ}$ and $P_{p^2kSZ}$ the
polarization signals induced by the tSZ and kSZ effects,
respectively. As shown by \cite{sunyaev1980}, these polarization
contributions depend only on the local density of free electrons and
not on their temperature. The corresponding amplitudes are
proportional to the square of the optical depth $\tau^2$
\citep{sazonov1999}, because of the two scattering events involved
into the process: consequently they are very sensitive probes of the
gas profile. The $P_{p^2kSZ}$ is one order of magnitude stronger than
the $P_{p^2tSZ}$, which can be even larger than the $P_Q$ signal,
depending on the considered frequency \citep{lavaux2004}.

\subsection{Possible contaminants}\label{sec:contamina_pol}

Many other processes can introduce significant contributions to the
linear polarization, like the Galactic synchrotron emission by our
Galaxy and by point radiosources, and the scattering with Galactic
dust grains. In general these processes have different frequency
dependences, allowing their separation through multi-band
observations, similarly to the techniques adopted for the CMB
temperature maps \citep[see, e.g.,][]{tegmark2000}.  The
component-separation method could also be improved by including in the
foreground subtraction our knowledge on the distributions of dust, gas
and magnetic fields in the Galaxy.  In particular, if one is mostly
interested in the polarization contributed by galaxy clusters, the
observations could be focussed in regions at high galactic latitudes,
where the foreground effect is minimal.  Also the polarization
originated by the CMB quadrupole ($P_Q$) and the one induced by the
thermal SZ effect ($P_{p^2tSZ}$) have a different frequency
dependence, favoring their separation.  In addition $P_Q$ has typical
scales larger than the kpSZ signal, so that the use of optimal spatial
filters \citep[see, e.g., ][]{hu2001,maturi2005} could drastically
reduce its contamination.

The magnetic fields, combined with the presence of free electrons in
galaxy clusters, may cause a Faraday rotation of the polarization
orientation, reducing its importance. This rotation angle is
relatively small also for massive galaxy clusters: $\Delta_{RM}\simeq
1-10^\circ (10 GHz/\nu)^2$ \citep{ohno2003}, however it could become
significant when considering our Galaxy at low galactic latitudes
\citep[see, e.g.,][]{burigana2006}.

It is well known that a polarization signal is also produced by both
the reionization process \citep[see, e.g.,][]{ng1996} and weak
gravitational lensing effect \citep[see, e.g.,][]{zaldarriaga1998}. In
particular the deflection of the CMB photons leads to a smoothing of
the small-scale features and to a power enhancement of the damping
tail in the power spectrum.  Notice that the analysis of the weak
lensing effect on the CMB has been suggested as possible tool to
reconstruct the cluster masses \citep{HU07.1}: in this case the kpSZ
signal represent a possible source of noise which must be taken under
control to avoid systematic effects.

In general all these contaminants should be analyzed in much more
detail, but the promises of observing the kpSZ signal in the next
future are interesting.

\section{Numerical simulations}\label{sec:pol_sim}

For the purpose of this work we use two different sets of resimulated
galaxy clusters.  Both have been extracted from a dark matter(DM)-only
simulation \citep[see][]{yoshida2001} of a ``concordance'' flat
$\Lambda CDM$ model, where the contributions to the density parameter
from matter and cosmological constant are $\Omega_{0m}=0.3$ and
$\Omega_{0\Lambda}=0.7$ respectively; the Hubble constant in units of
100 km s$^{-1}$ Mpc$^{-1}$ is $h=0.7$.  The initial conditions of the
parent simulation, that followed the evolution of $512^3$ particles
with mass of $6.86 \times 10^{10}$ $h^{-1} M_\odot$ in a comoving
cubic box of $479\,h^{-1} \,\mbox{Mpc}$ per side, were set considering
a cold dark matter power spectrum normalized by assuming
$\sigma_8=0.9$.

In the catalogue of haloes identified in final output of the parent
simulation by adopting a spherical overdensity criterion, we selected
few objects to be re-simulated with higher mass and force resolution
The new initial conditions have been created by using the ``Zoomed
Initial Conditions'' (ZIC) technique \citep{tormen1997}. This method
identifies in the initial domain the Lagrangian region corresponding
to each cluster up to few virial radii and populates it with more
particles (of both DM and gas), properly adding small-scale modes. At
the same time the volume outside the region of interest is sampled by
low-resolution particles which allow to reproduce the original tidal
effects of the cosmological environment.

The new re-simulations, starting at redshift $z_{\rm in}$ ranging
between 35 and 60, have been carried out using for the first sample
the Tree-Smoothed Particle Hydrodynamics (SPH) parallel code GADGET
\citep{springel2001}, while for the second set we ran its new version
GADGET-2 \citep{springel2005}, which includes an entropy-conserving
formulation of SPH \citep{springel2002} and allow, by choice, the
treatment of many physical processes affecting the baryonic component
(see below).  The gravitational softening length was kept fixed at
$30\,h^{-1} \,\mbox{kpc}$ comoving (Plummer-equivalent) for $z>5$, and
was switched to a physical softening length of $5\,h^{-1}
\,\mbox{kpc}$ at smaller redshifts.

The first sample comprises 17 randomly selected galaxy clusters, whose
structure and dynamical properties has been discussed in
\cite{rasia2004} and \cite{tormen2004}, respectively.  These
simulations, which follow only non-radiative hydrodynamics, assume a
baryonic density corresponding to $\Omega_{0b}=0.03$; the particles
masses in the high-resolution region range from $2 \times 10^9
\,h^{-1} M_\odot$ to $6 \times 10^9 \,h^{-1} M_\odot$ for DM, and from
$3 \times 10^8 \,h^{-1} M_\odot$ to $7 \times 10^8 \,h^{-1} M_\odot$
for gas.  The final objects have virial masses between $3.1 \times
10^{14} \,h^{-1} M_\odot$ and $1.7 \times 10^{15}
\,h^{-1} M_\odot$, and virial radii\footnote{Virial radii are
defined using the overdensity threshold dictated by the spherical
top-hat model \citep[see, e.g.,][]{eke1996}.} between 1.4 and $2.5
\,h^{-1}\,\mbox{Mpc}$.

The second sample comprises 9 more isolated objects, 4 of them with
the typical size of clusters (about $10^{15} \,h^{-1} M_\odot$), and 5
of group-like systems (about $10^{14} \,h^{-1} M_\odot$).  In this
case the baryonic density has been fixed to $\Omega_{0b}=0.04$.  Each
cluster of this set has been re-simulated 4 times, turning on
different kinds of physical processes:
\begin{enumerate}
\item
gravitational heating only with an ordinary treatment of gas viscosity
(hereafter {\it ovisc} model);
\item
like {\it ovisc}, but using an alternative implementation of the
artificial viscosity (hereafter {\it lvisc} model), which allows to
better resolve the turbulence driven by fluid instabilities: this
leads to a non-negligible contribution of non-thermal pressure support
in the central region of galaxy clusters \citep{dolag2005};
\item
cooling, star formation and supernovae feedback with weak galactic
winds, having a speed of $\sim 340$ km/s ({\it csf} model);
\item
like {\it csf}, but also including thermal conduction
\citep[see][]{jubelgas2004,dolag2004}, where the isotropic effective
conductivity is fixed to 1/3 of the Spitzer rate ({\it csfc} model).
\end{enumerate}

A more detailed presentation of the properties of the clusters of this
second set can be found elsewhere \citep[see,
e.g.,][]{ettori2006,roncarelli2006}.

We want to stress here that the simulated clusters of the first sample
have been randomly chosen from the complete list of haloes identified
at $z=0$ in the parent simulation, with the only requirement of
sampling in a sufficiently complete way the mass range above
$10^{14}\,h^{-1} M_\odot$.  For this reason they have varying
dynamical and environmental properties, comprising relaxed and
non-relaxed objects, isolated and interacting systems.  This allows us
to consider this sample as representative of the whole population of
galaxy clusters.  On the other hand the clusters of the second set
have been chosen to be more isolated structures, so that their main
properties are not affected by recent major merger events and thus the
effects of the four different ICM models can be better identified.

To compute the final Stokes parameters related to the kpSZ signal, we
need to integrate Eq.~(\ref{eqn:dQ_dU}) along the line-of-sight. This
is done by considering boxes of $(8\,\mbox{Mpc})^3$ centered on the
cluster: the resulting maps have a resolution of $15.6\,\mbox{kpc}$.
The polarization contributed by each particle, or moving cloud, adds
constructively if their tangential velocities have the same direction,
even if they have opposite versus. On the other hand, if their
tangential velocities are orthogonal, the polarization tends to
cancel. Because of that, only the coherent motion of gas clouds
produces a polarization contribution, unlike the thermal motion of
electrons, which is randomly oriented and does not contribute to the
signal.

\section{The importance of the merging activity}\label{sec:examples}

\subsection{Two examples with different dynamical state}

In order to discuss the role of the merging activity, we analyzed in
more detail two clusters of our first sample, named g24+200 and
g7. The first object represents a relaxed halo falling with high
velocity into the potential well of another cluster. Consequently the
velocity field of its gas is dominated by the bulk motion of the whole
system. For this reason it can be considered at the same time like an
example of a relaxed halo and of a major merger event because of its
bulk motion given by the companion cluster. Conversely, the second
object is not relaxed and is acquiring new matter from different
in-falling subclumps having high velocities.

\begin{figure*}[!t]
  \centering
  \includegraphics[width=15cm]{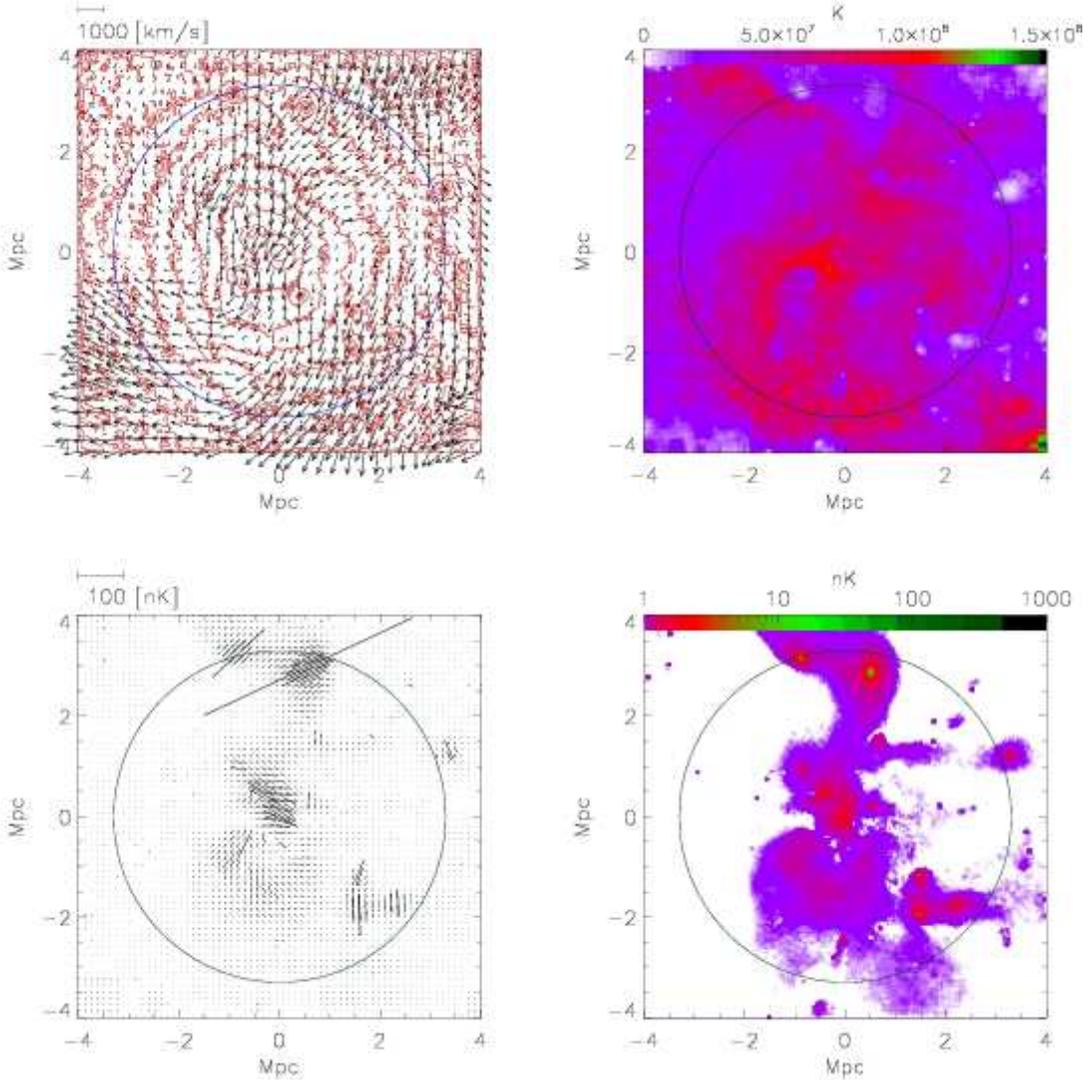}
  \caption{
Maps of a region of $8 \times 8 \,\mbox{Mpc}^2$ centered on the
simulated cluster g7.  Upper left panel: arrows represent the
(mass-weighted) tangential velocity of the gas; the overlaid levels
refer to the density iso-contours.  Upper right panel: the gas
(mass-weighted) temperature.  Bottom left panel: the direction of the
polarization pseudo-vector field.  Bottom right panel: the polarized
radiation amplitude.  The different scale units or color levels are
shown above the corresponding panels.  In all panels, the circle
indicates the virial radius.  }
\label{fig:g7}
\end{figure*}

Figure~\ref{fig:g7} shows a region of $8 \times 8 \,\mbox{Mpc}^2$
centered on cluster g7. This object has a virial mass of $1.46 \times
10^{15} \,h^{-1} M_\odot$ and a virial radius of $2.35
\,h^{-1}\,\mbox{Mpc}$ (indicated by the circle in the plots).  In the
upper-left panel we overlay the mass-weighted tangential velocity
field of the gas to its density iso-contours (red curves); the arrow
length is proportional to the velocities. The upper-right panel shows
the corresponding map for the mass-weighted temperature of the gas;
the color scale is given in the upper part of the panel. In the bottom
left-panel we show the polarization pseudo-vector field, where the
lengths of the lines are proportional to the polarization
amplitude. Note that, for graphical purposes, the lines sample the
signal on a grid having a lower resolution, giving only an idea of the
polarization and velocity directions. Finally, we draw in the
bottom-right panel the amplitude of the polarization signal using a
logarithmic scale (reported in the upper part). The polarization maps
are computed at a frequency of $300 \,\mbox{GHz}$.

Since the polarization is proportional to the projected density and to
the squared tangential velocity (see Eq.~\ref{eqn:dQ_dU}), the signal
is peaked on the core of the merging haloes, where the column density
of electrons and the tangential bulk motion are large. The kpSZ signal
of some in-falling blobs can be even stronger than the cluster core
itself: an example is the blob approximately located at the position
($0.5 \,\mbox{Mpc}$ , $3.0 \,\mbox{Mpc}$).

\begin{figure*}[!t]
  \centering
  \includegraphics[width=14cm]{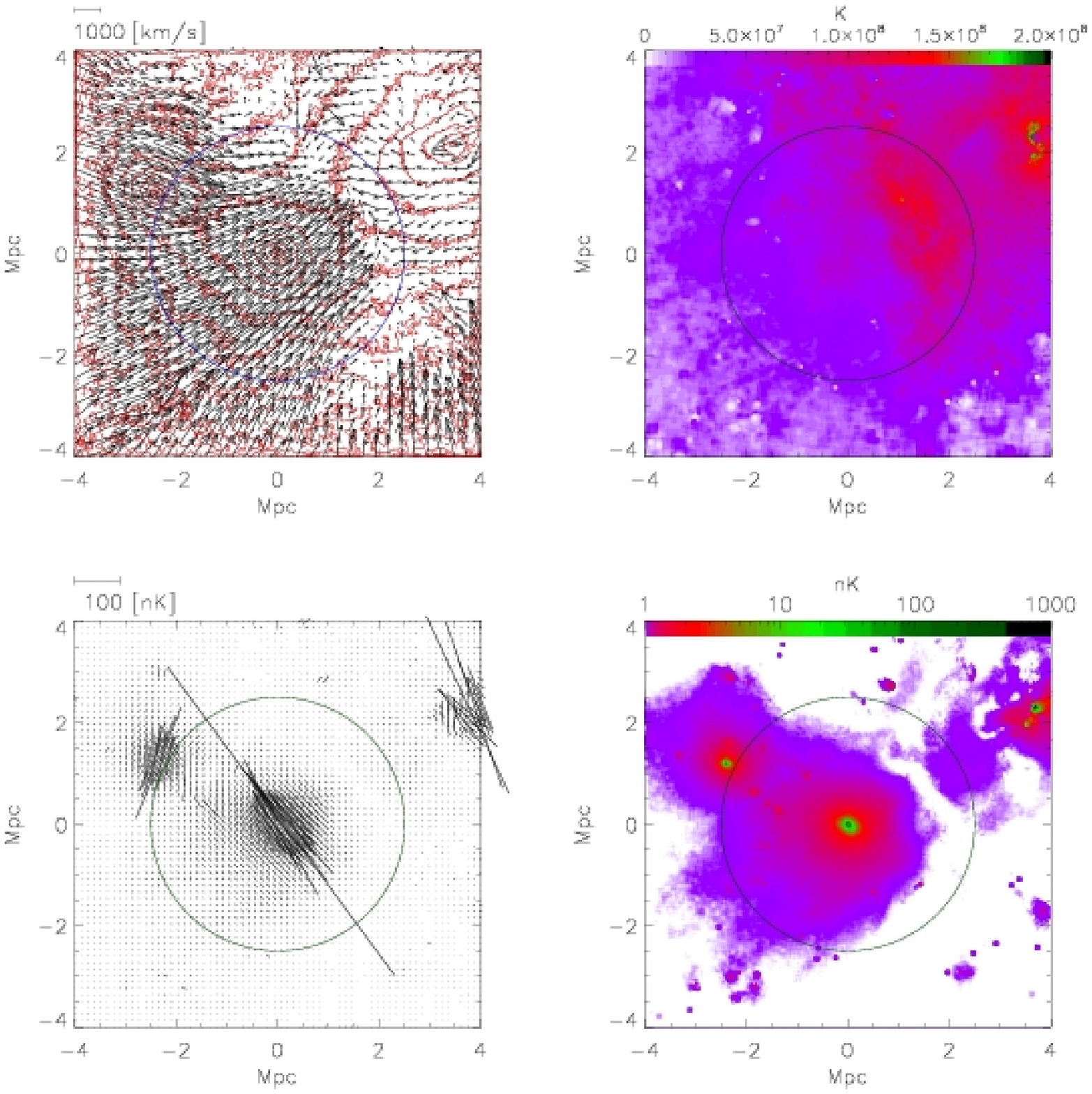}
  \caption{
  As Figure~(\ref{fig:g7}), but for the simulated galaxy cluster g24+200.}
  \label{fig:g24+200}
\end{figure*}

In Figure~\ref{fig:g24+200} we display the same quantities for the
cluster g24+200, which has a virial mass of $6.54 \times 10^{14}
\,h^{-1} M_\odot$ and a virial radius of $1.80 \,h^{-1}\,\mbox{Mpc}$.
Unlike cluster g7, for this object the polarization inside the virial
radius is dominated by the cluster core where the signal shows a
constant direction.  Outside the virial radius there are two more
evident structures which do not belong to this cluster: the first one
is a small but fastly in-falling blob, the second one, located at the
position ($3.8 \,\mbox{Mpc}$, $2.4 \,\mbox{Mpc}$), represents a
different cluster.  The absolute value of the polarization signal of
this cluster is large because it is taking part of a major merger
event: in fact it is moving with a bulk velocity of
$1200\,\mbox{km/s}$ into the direction of the other cluster visible in
this field.

We discuss the differences between these two clusters and the
importance of their merging activity into the following subsections.

\subsection{Excluding the merging activity}\label{sec:no_merg}

In order to better understand the influence of the merging activity on
the polarization signal, we compare the polarization map displayed in
Figs.~\ref{fig:g7} and \ref{fig:g24+200} with the ones obtained by
artificially ``freezing'' the internal dynamics. This is done by
associating the mean bulk velocity\footnote{In order to avoid the
contamination given by in-falling blobs, this is defined as the
velocity associated with the baricentre of the most inner 50 per cent
of the cluster mass.} of the whole cluster to all particles inside the
box, so that the cluster moves like a rigid body. For simplicity
hereafter we will refer to these simulations as {\it ``merging-free''
simulations}.

\begin{figure*}[!t]
  \centering \includegraphics[width=14cm]{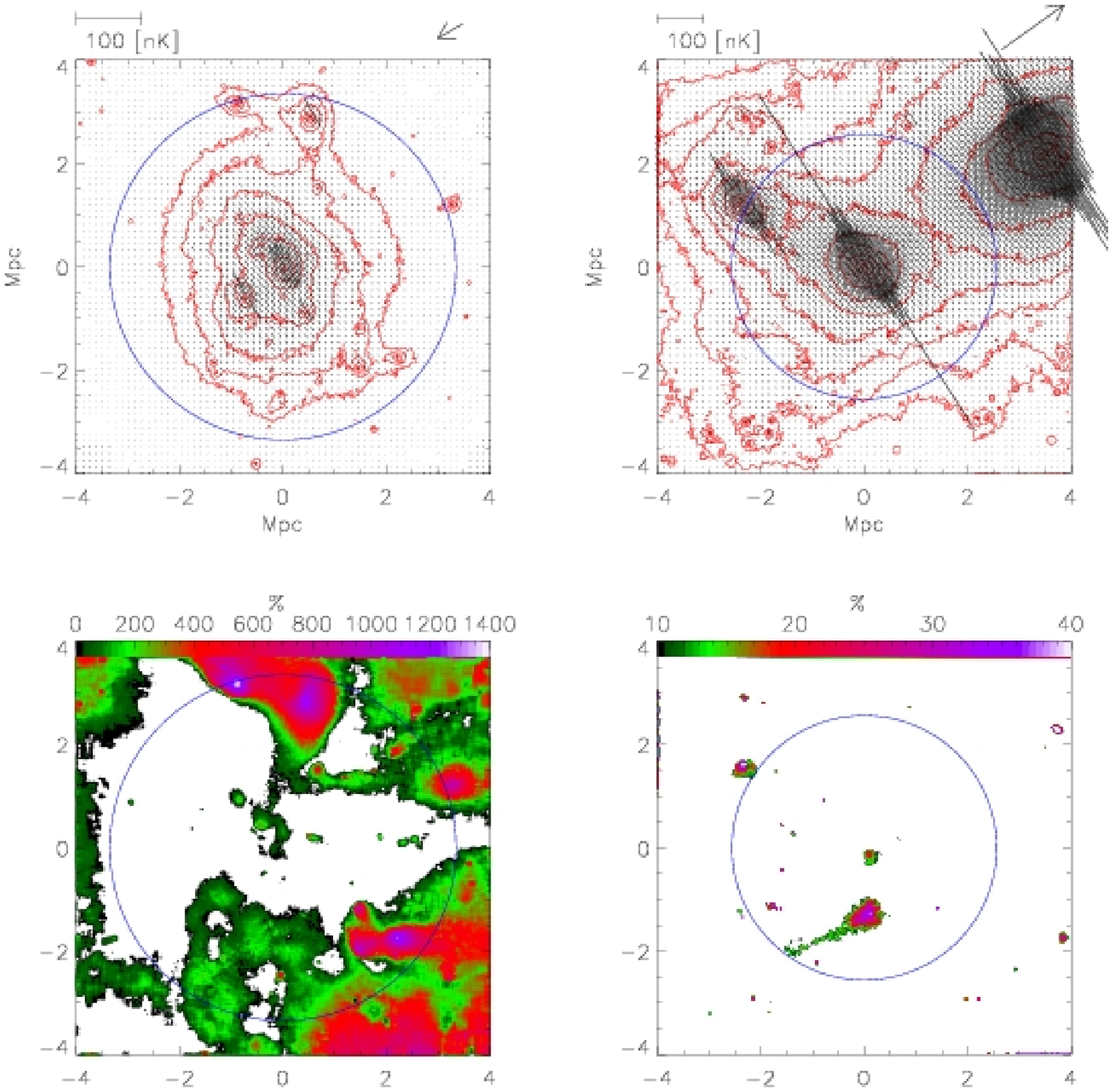}
  \caption{Comparison with ``merging-free'' simulations.  Left and
    right panels refer to the dynamically active cluster g7 and to the
    more relaxed cluster g24+200, respectively.  Top panels: the
    polarization signal obtained from the ``merging-free''
    simulations, shown by adopting the same scales and units of
    Figs. \ref{fig:g7} and \ref{fig:g24+200}. The iso-polarization contours
    start from $1\,\mbox{nK}$ and show levels 
    power of 2 steps. An arrow proportional to the bulk velocity of each
    cluster and having its direction is displayed on the top right
    part. Bottom panels: maps for the per cent increment of the
    polarization signal, given by the merging activity, as expressed
    by Equation~(\ref{eqn:increment}). The corresponding color scale
    is shown above the panels.} \label{fig:increment}
\end{figure*}

The top panels in Figure~\ref{fig:increment} show the polarization
signal obtained in such a way, for the unrelaxed cluster g7 (on the
left) and for the relaxed cluster g24+200 (on the right).  The
polarization signal is proportional to the length of the lines and it
is displayed using the same scale adopted for Figs.~\ref{fig:g7} and
\ref{fig:g24+200}. On the top of these panels, we also report an arrow
corresponding to the bulk velocity of each galaxy cluster. The
iso-polarization contours in the upper panels start from
$1\,\mbox{nK}$ and show levels in power of 2 steps, i.e
$2,4,8,16\,\mbox{nK}$, etc.  Note that the large structure in the
upper-right corner of the g24+200 panel is an artifact due to the
procedure performed to obtain the ``merging-free'' simulation. In
fact, the bulk motion of the cluster could be assigned only to the
particles inside the virial radius.  Consequently, in the
``merging-free'' simulations only the region inside the virial radius
has a physical meaning and will be considered in the following
discussion.

\subsection{Quantifying the importance of the merging}\label{sec:importance}

As described above, we computed the kpSZ signal of our two clusters,
considering both for the real simulations and the ``merging-free''
versions of them.  Comparing the corresponding results, we can
evaluate the amplification of the polarization signal produced by the
merging activity.  This is done by computing the per cent increment of
the signal, defined as
\be
\label{eqn:increment}
\Delta (\vec{\theta}) \equiv \frac{P(\vec{\theta})
                                      - P_{\rm mf}(\vec{\theta})}
                                        {P_{\rm mf}(\vec{\theta})}  \,,
\ee
where $P$ and $P_{\rm mf}$ are the polarization signals given by the
real and ``merging-free'' simulations, respectively.  A map of this
quantity is shown for both clusters in the bottom panels of
Figure~\ref{fig:increment}.  Clearly the relaxed cluster in the left
panel strongly differs from the non-relaxed object in the right panel.
In the first case (g24+200) there is only a small increment of the
signal (30 per cent at most) and only over a small area centered on a
small halo; the signal of the cluster core is substantially unchanged,
showing an increment by only 10 per cent.  In the second case (cluster
g7) a large fraction of the object shows a substantial signal
increment, especially in the outer regions, where most of the
in-falling blobs are located.  Here the merging activity produces an
increment by 170 per cent in the cluster core and up to a factor of 10
at the positions of five different sub-halos, three of which have a
signal even stronger than the cluster core itself.

\begin{figure*}[!t]
  \centering 
  \includegraphics[width=7cm]{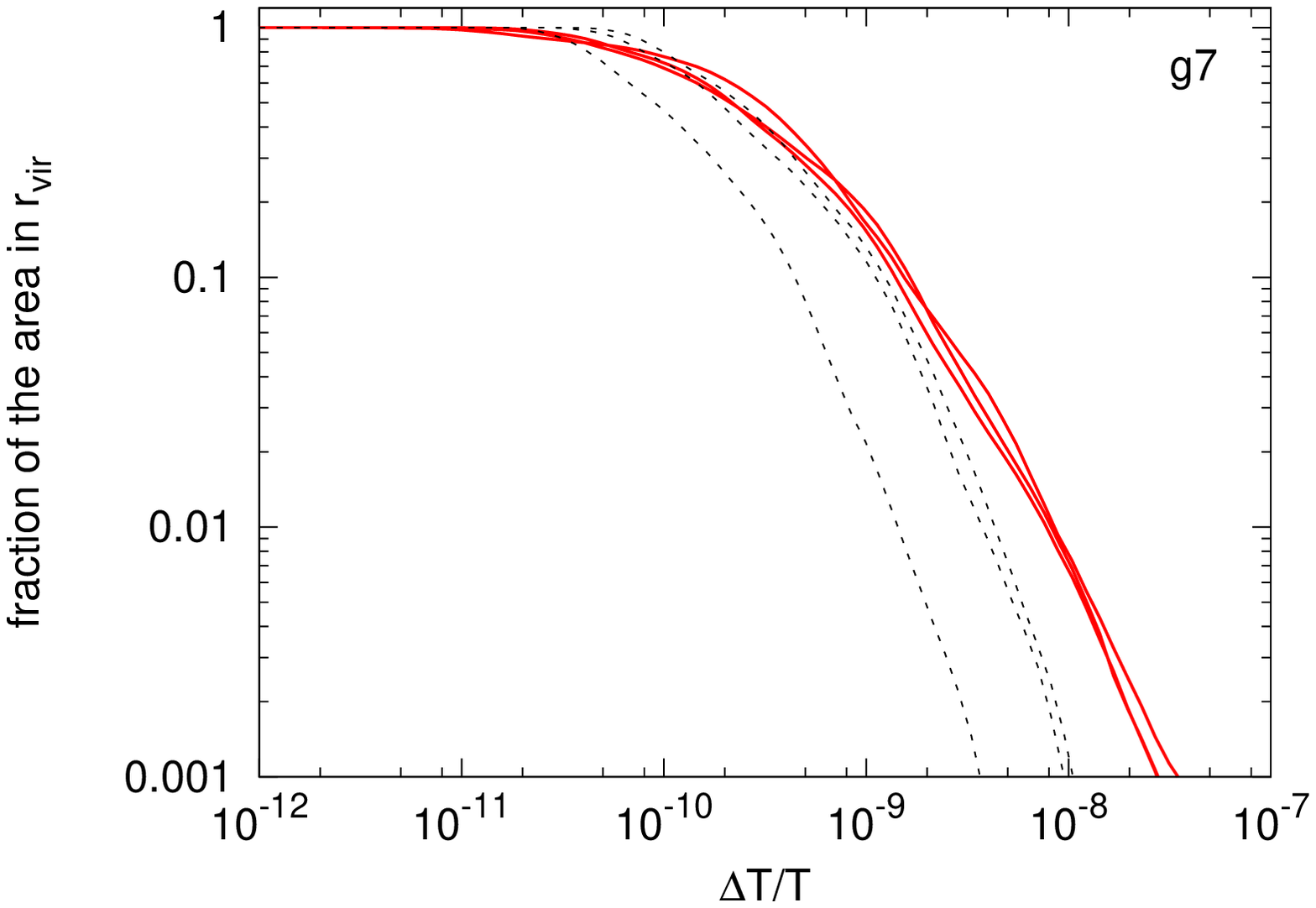}
  \includegraphics[width=7cm]{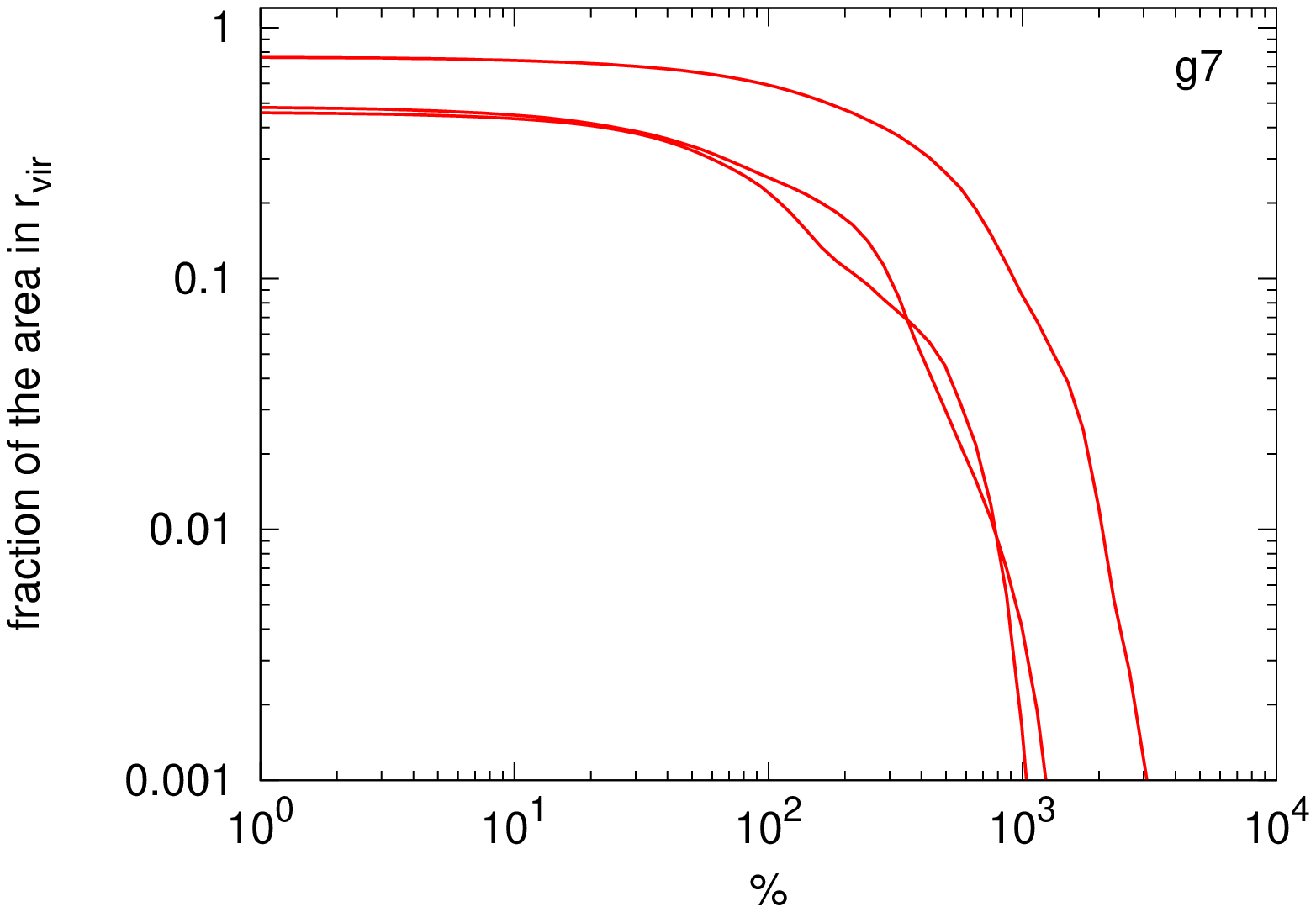}
  \includegraphics[width=7cm]{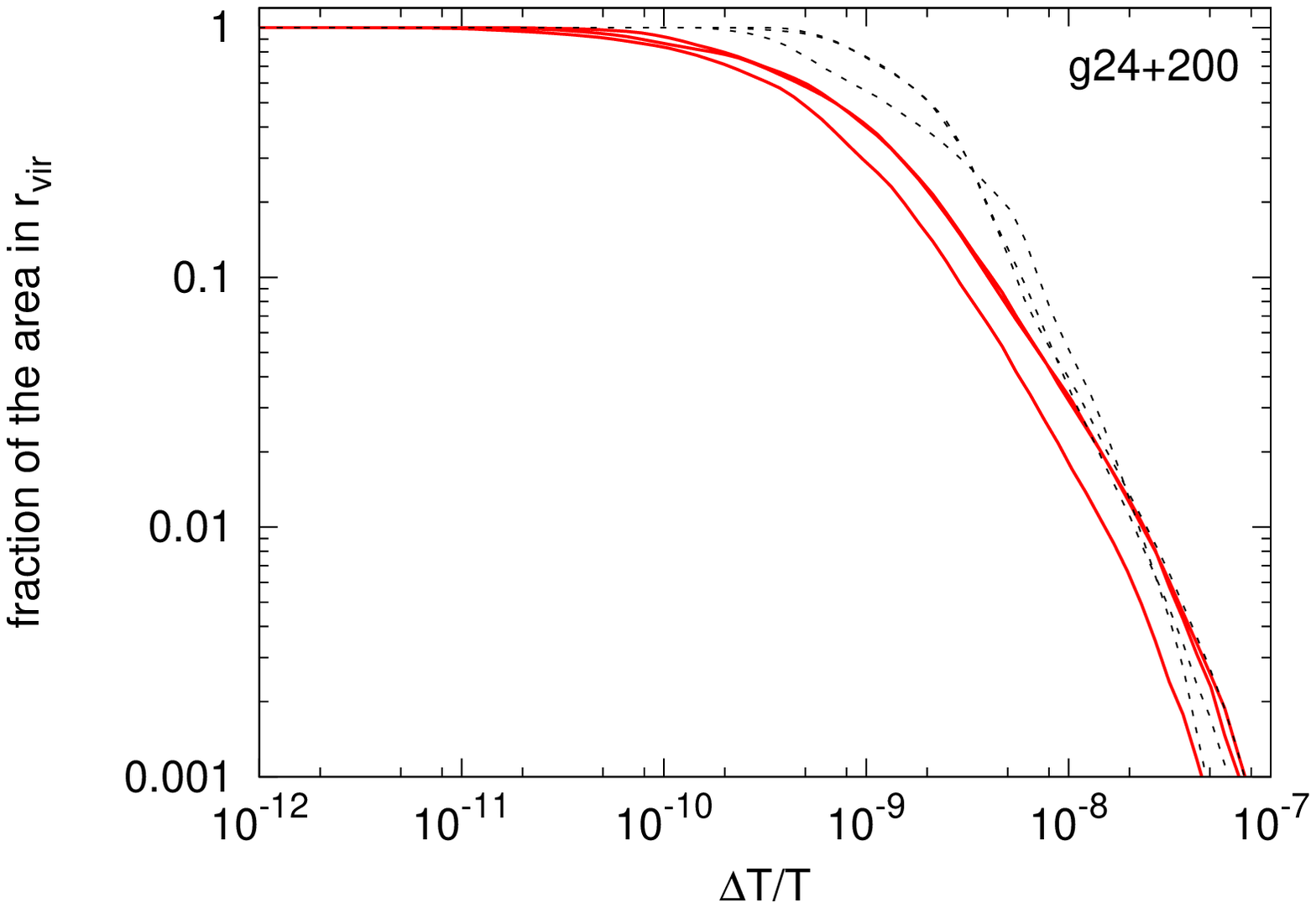}
  \includegraphics[width=7cm]{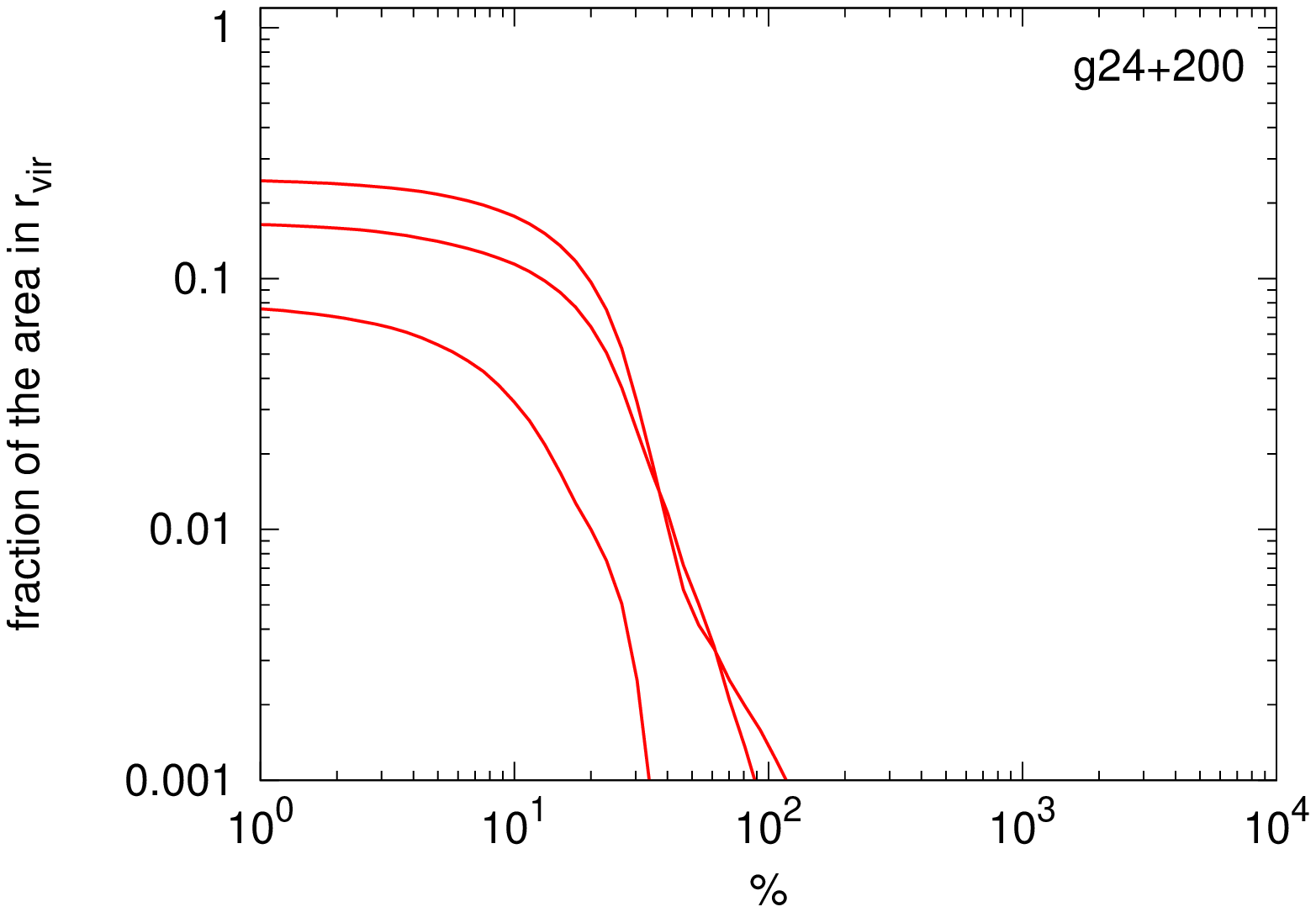}
  \caption{ The statistical distribution of the polarization signal
    for the merging cluster g7 (top panels) and for the relaxed
    cluster g24+200 (bottom panels). The results for three different
    cartesian projections of each simulated cluster are shown.  The
    left panels present the fraction of the virial area covered by the
    cluster having a polarization signal larger than a given threshold
    (expressed in terms of $\Delta T/T$); solid and dotted lines refer
    to results for the original simulations and for the `merging-free'
    simulations, respectively.  The right panels show the area
    fraction where the per cent variation of the polarization amplification
    produced by the merging activity (defined as in
    Equation~\ref{eqn:increment}) is larger than a given threshold.  }
\label{fig:cum_g7_g24}
\end{figure*}

In order to better describe the distribution of the polarization, in
Figure~\ref{fig:cum_g7_g24} we report the fraction of the virial area
covered by the galaxy clusters where the signal is larger than a given
threshold.  The results for the merging cluster g7 and for the relaxed
system g24+200 are shown in the upper and bottom panels, respectively;
in this case all three orthogonal projections have been considered.
The solid lines refer to the original simulations (i.e. including the
merging activity), while the dotted lines correspond to the
``merging-free'' simulations.  A more quantitative evaluation of the
differences between relaxed and un-relaxed clusters can be obtained by
looking to the right panels, where we plot the fraction of the virial
area as a function of the variation of the polarization signal
produced by the merging activity.  It is evident that for the relaxed
object g24+200 there is not any significant polarization amplification
from the inner dynamics: only $1$ per cent of the cluster area shows
an amplification of at least $30$ per cent.  On the contrary, the
un-relaxed cluster g7 shows an increment of at least $300$ per cent
(800 per cent) on $10$ per cent (1 per cent) of the cluster area; 50
per cent of the cluster area inside the cluster has an amplification
of 50 per cent.  Note that the distributions in the right panels do
not reach the unity, because we considered only positive variations.
Thus the difference between unity and the maximum value represents the
fraction of the area within the virial radius with a negative
$\Delta$: even if it may be large, we checked that it corresponds to
regions where the polarization signal is negligible.

We notice that one of the three projections of cluster g7 has a
significantly lower bulk velocity with respect to the other ones (379
against 577 and $624\,\mbox{km/s}$). As a consequence, in the
merging-free simulation for this particular projection the
polarization signal is lower with respect to the other ones, but as
high as the other cases when the merging activity is fully included.
This shows that the gas dynamics originated during the merging events
dominates over the cluster bulk motion.

\begin{table}[!t]
  \caption{Per cent variation of the polarization signal for clusters
  g7 and g24+200 placed at redshift $z=0.1$. The table reports the
  values corresponding to two different area fractions (10 and 1 per
  cent) and to four different beam resolutions (ranging from 0 to 10 arcmin).}
  \label{tab:cum_summary}
  \begin{center}
    \begin{tabular}{|c|c|c||c|c|}
      \hline
      beam &  g7 (10\%)  &  g7 (1\%)  & g24+200 (10\%) & g24+200 (1\%)\\
      \hline
      0'   &  550        & 1500       & 10             & 35      \\
      1'   &  490        & 1500       & 10             & 30      \\
      5'   &  480        & 1400       & 7              & 22      \\
      10'  &  477        & 1044       & 0.5            & 14      \\
      \hline 
    \end{tabular}
  \end{center}
\end{table}

It is important to study how the previous results change considering
different instrumental resolutions. This is evaluated by placing the
galaxy clusters at redshift $z=0.1$ and convolving the original maps
with four different aperture beams.  Table~
\ref{tab:cum_summary} summarizes the results obtained considering all
three projections.  Since the polarization signal contributed by the
in-falling blobs is peaked on small scales, the convolution strongly
smooths down the amplification produced by the merging activity.
Obviously this effect is larger for lower resolutions, but it affects
only the high-polarization region, leaving unchanged the signal on
larger scales.

Concluding this section, we notice that the example of these two
clusters clearly shows the importance of the merging activity for the
polarization signal and suggests two main ways to exploit this
amplification to extract useful information. First, polarization could
be used to investigate the ICM dynamics. In fact it could be possible
in principle to obtain a complete three-dimensional reconstruction of
the velocity field thanks to the combination with kSZ measurements
\citep{nagai2003}: this would provide a picture of the merging history
of our universe.  Second, the kpSZ signal could be used to determine
the tangential bulk motion of clusters, allowing a determination of
the three-dimensional velocity power spectrum (always in combination
with kSZ data). This would lead to independent and complementary
estimates of cosmological parameters \citep[see,
e.g.,][]{tormen1993,moscardini1996,bhattacharya2007}.  In this case
the merging activity represents a strong source of noise and a
particular care should be used in selecting clusters.  To discuss in
more detail the reliability of these two possibilities, we need to
repeat the previous analysis on more robust statistical bases. For
this reason we analyze the entire samples of simulated clusters in the
next section.

\section{Analysis of the full cluster samples}\label{sec:statistic}

\subsection{Non-radiative simulations}

First, we perform a statistical analysis considering our first sample,
composed by 17 different haloes simulated under the assumption of
non-radiative hydrodynamics.  For all objects, we consider the
projections along three orthogonal lines-of-sights: this leads to 51
polarization maps. For simplicity we placed all systems at redshift
$z=0.1$.

To investigate the effect of the instrument resolution, we convolve
all polarization maps with a Gaussian kernel assuming different
aperture beams. This is of primary importance because the largest
contributions to the total signal come from small haloes, which could
be smeared out by a too large beam.

\begin{figure*}[!t]
  \centering
  \includegraphics[width=7cm]{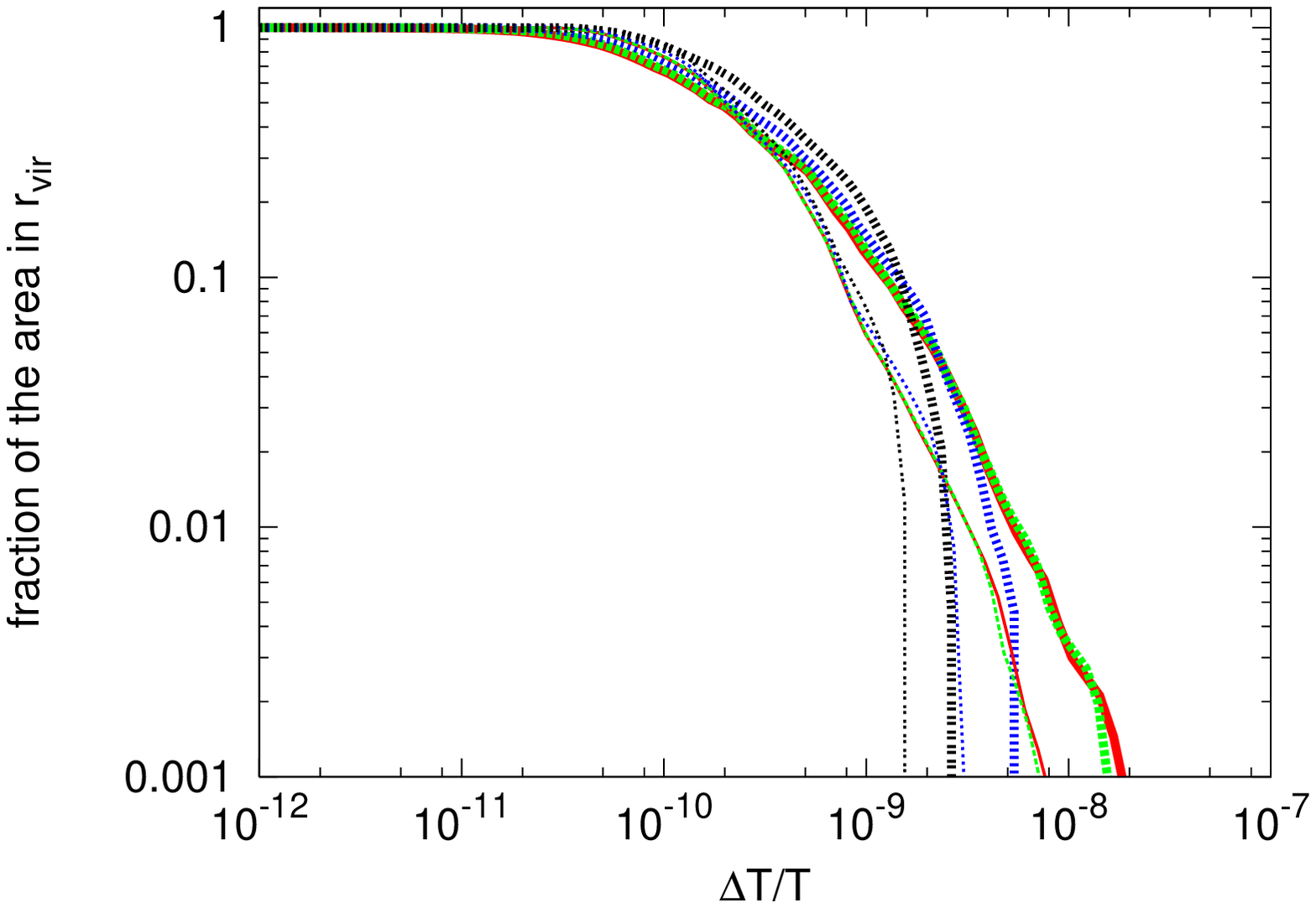}
  \includegraphics[width=7cm]{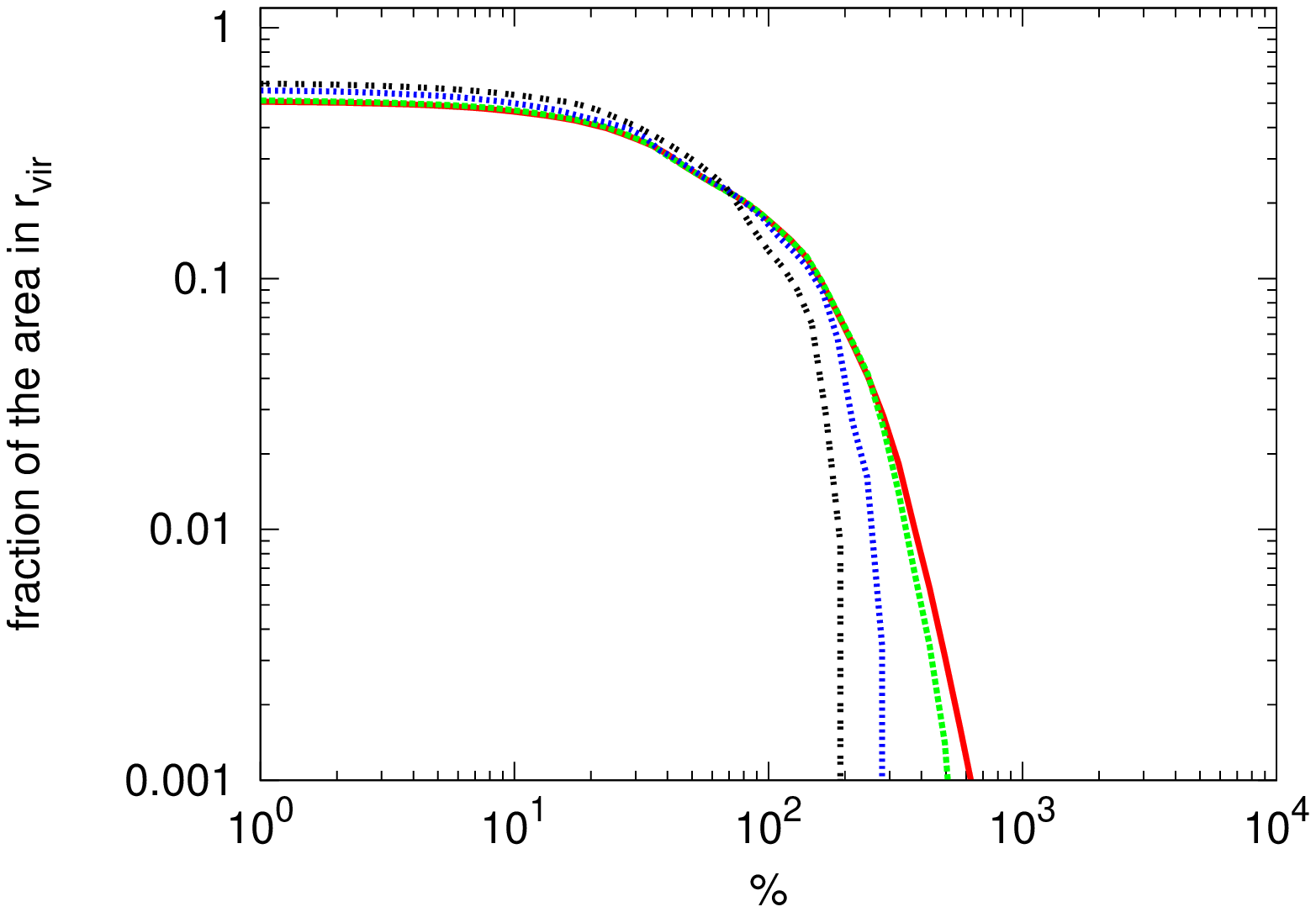}
  \caption{ The statistical distribution of the polarization signal as
  derived from the analysis of all projections of the 17 simulated
  galaxy clusters belonging to our first sample.  Left panel: the
  fraction of the area having a polarization signal larger than a
  given threshold (expressed in terms of $\Delta T/T$); thick and
  thin lines refer to results for the original simulations and for
  the `merging-free' simulations, respectively.  Right panel: the area
  fraction where the per cent amplification of the polarization signal
  produced by the merging activity is larger than a given
  threshold. Curves of different colors present the results for the
  original maps (red lines) and for different aperture beams: 1'
  (green line), 5' (blue line) and 10' (black line). 
}
\label{fig:cum_statistica}
\end{figure*}

In the left panel of Figure~ \ref{fig:cum_statistica} we plot the
cumulative distribution of the kpSZ signal as a function of the
covered area inside the virial radius. The different curves represent
the median of the 51 realizations and present the results obtained
from the original maps (red line) and for different aperture beams: 1'
(green line), 5' (blue line) and 10' (black line).  The thick lines
refer to the simulations including the merging activity, while the
thin lines are the results for the ``merging-free'' simulations,
i.e. where only the cluster bulk motion is accounted for. Using the
same line coding, in the right panel of
Figure~\ref{fig:cum_statistica} we show the distribution of per cent
amplifications for the polarization signal originated by the merging
activity. As expected, the instrumental resolution causes a
significant reduction of the maximum amplitude of the signal,
decreasing also the increment induced by merging.  The minimum
aperture beam affecting the signal is obviously related to the typical
scale of the objects producing the polarization peaks. This scale is
$\sim 1$ arcmin which, not surprisingly, is of the same order of the
halo core radius ($r_c\sim 0.11 h^{-1}Mpc$ at $z=0.1$).

Our results suggest that the upcoming millimetric observatories,
having angular resolutions of the order of $FWHM\simeq 1'$ or better,
will be sensitive to the merging activity of galaxy clusters, which is
expected to amplify the kpSZ signal by a factor of about 5 on at least
1 per cent of the virial area.  The plot also shows that on average
the cluster population has a maximum signal of $\Delta T/T\approx
10^{-8}$, while merging systems have peaks up to $10^{-7}$: this
confirms what we obtained from the analysis of two clusters g7 and
g24+200 in the previous section.

We notice that this median distribution, which represents the typical
expectation for the composite cluster population including both
relaxed and non-relaxed objects, is consistent with the results
obtained by \cite{lavaux2004}. However we point out that the merging
clusters may have a kpSZ signal 10 times stronger.

Additionally to the dynamic information carried on by this effect, it
is important to notice that the kpSZ effect can become a
non-negligible source of noise in the estimate of the CMB quadrupole
when observing the polarization induced by the intrinsic CMB
quadrupole $P_Q$.  This makes necessary both to have access to a
component-separation technique based on multi-frequency observations
and to carefully choose the clusters to be considered, which should be
as relaxed as possible.  This selection can be performed thanks to
X-ray and SZ observations.

\subsection{Effects of the ICM modeling}\label{sec:ICMmodelling}

Finally, we analyze the second simulation set to investigate the
effects of different ICM modeling onto the statistical properties of
the kpSZ signal. Again, the 9 objects have been projected onto 3
different orthogonal directions, leading to 27 polarization maps for
each of the ICM modeling described in Sect.~\ref{sec:pol_sim}.

\begin{figure*}[!t]
  \centering
  \includegraphics[width=14cm]{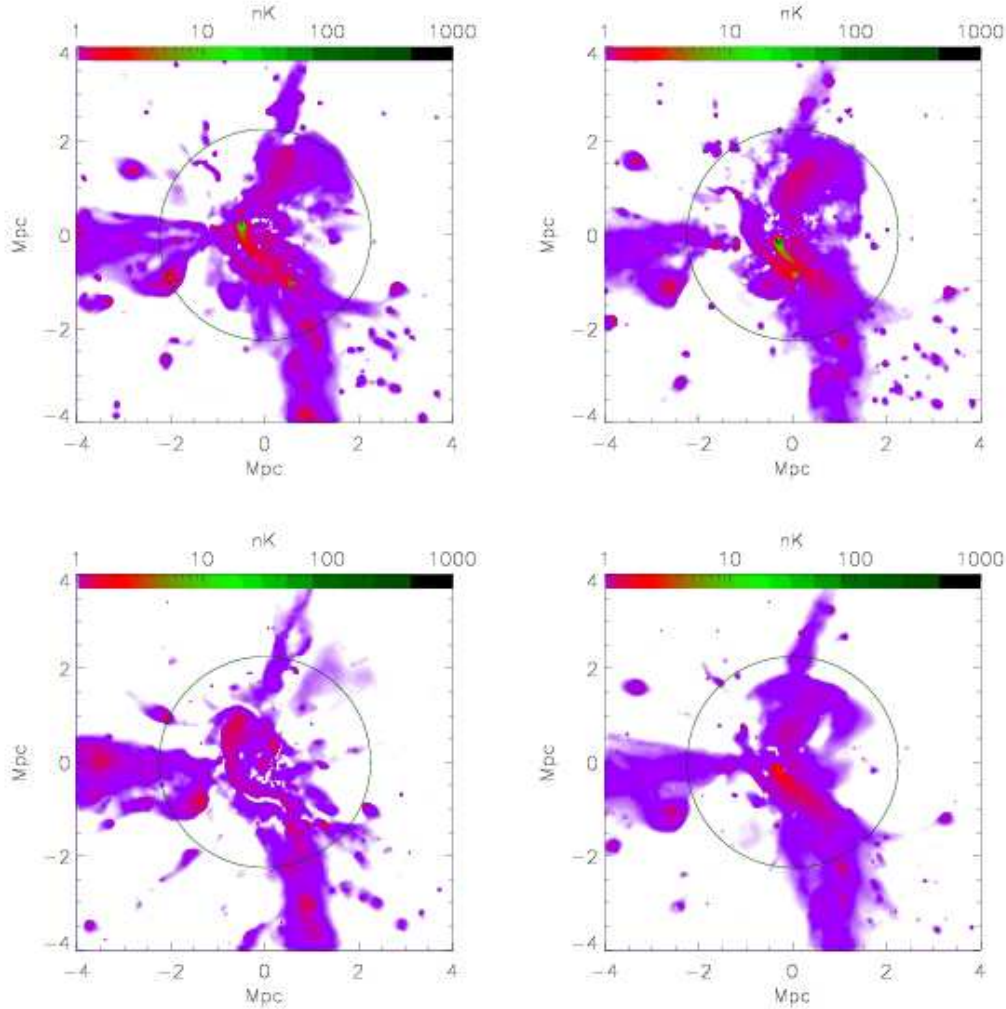}
  \caption{Maps for the polarization signal for the same cluster, g51,
  but simulated with different ICM modeling: {\it ovisc} (top left),
  {\it lvisc} (top right), {\it csf} (bottom left), {\it csfc}
  (bottom right).  The same projection is shown in all panels; the
  circle represents the virial radius.} \label{fig:g51_phys}
\end{figure*}

As an example, we shown in Figure \ref{fig:g51_phys} the polarization
maps for the unrelaxed cluster g51 as obtained using the same
projection, but with the inclusion of different physical processes.
This object has a mass of $\approx 1.3\times 10^{15}\ h^{-1} M_\odot$
and a virial radius of $\approx 2.3\ h^{-1}$ Mpc (corresponding to the
circle in the plot).  The properties of the large-scale polarization
signal, coming from diffuse structures, are similar in the different
panels, reaching values of the order of $\sim 1 nk$. We find
significant changes only in the most central regions of the cluster,
where the different ICM modeling plays an important role.  In
particular, from the comparison of the two upper panels, it is evident
that the {\it ovisc} and {\it lvisc} models give almost equivalent
kpSZ signals, with consistent maximum amplitudes ($967\,nk$ and
$1123\,nk$, respectively).  Also the different features present in the
map have coincident positions and very similar appearence (they are
only slightly smoother in the {\it ovisc} model).  Since the only
difference in the ICM modeling implemented in these two simulations is
the possibility of better resolving the turbulent motions driven by
fluid instabilities in the {\it lvisc} model, our results seem to
suggest that their relevance for polarization is negligible.  The
processes of cooling, star formation and supernovae feedback with weak
winds which are active in the {\it csf} simulation, produce the effect
of smearing out the highest peaks in the cluster core, strongly
reducing the maximum signal ($285\,nk$), almost half order of
magnitude smaller than in the previous cases.  The general properties
of the polarization signal in the {\it csfc} model are similar to the
{\it csf} case, with the exception at the small scales, where the
signal is slightly smoothed down by the effect of the thermal
conduction.

\begin{figure*}[!t]
  \centering
  \includegraphics[width=14cm]{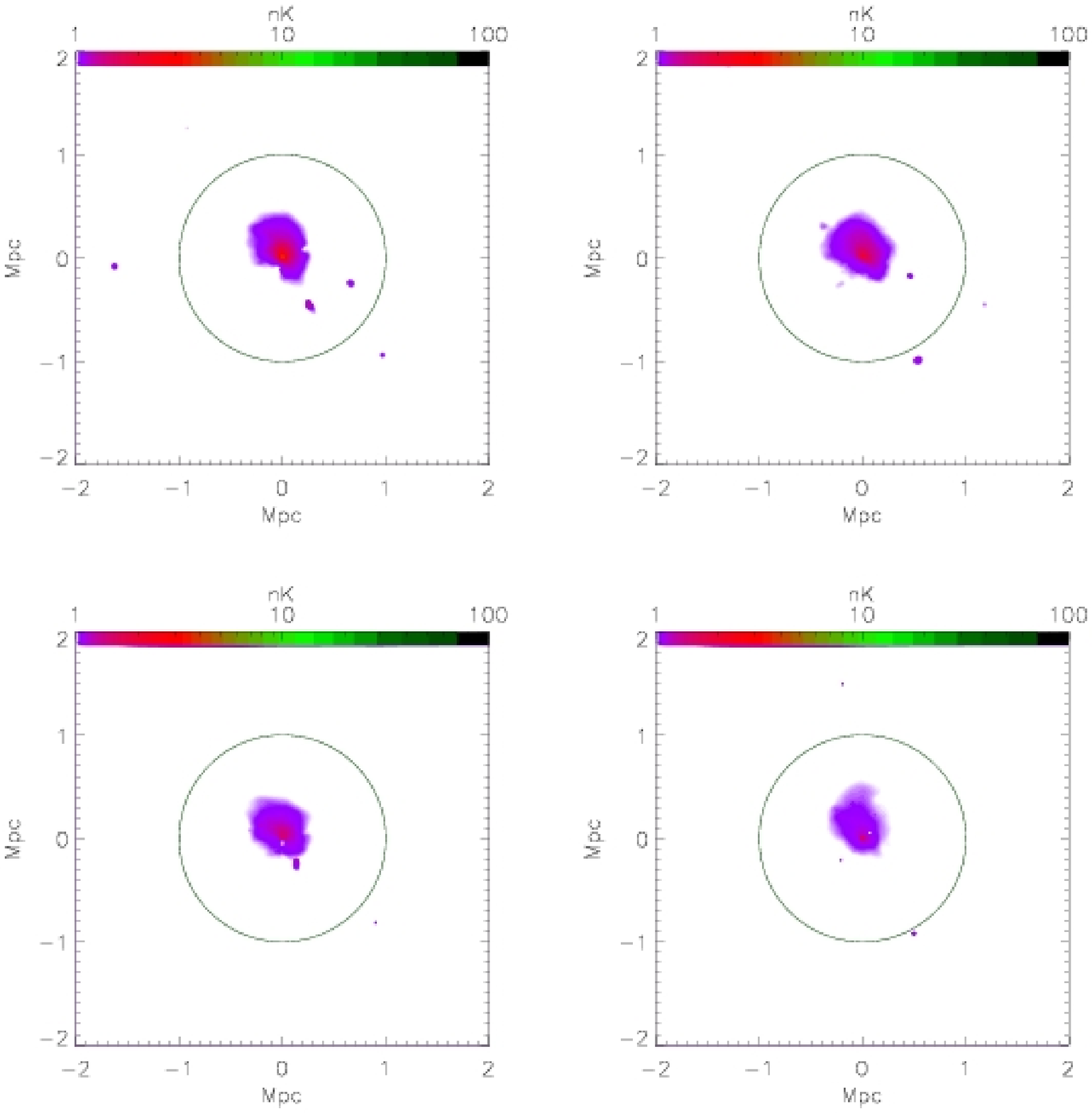}
  \caption{As Figure~ \ref{fig:g51_phys}, but for the more relaxed and
  smaller cluster g6212. }
  \label{fig:g6212_phys}
\end{figure*}

In Figure \ref{fig:g6212_phys} we show the polarization maps, but for
the more relaxed and smaller cluster g6212 (having virial mass and
radius of $\approx 1.15\times 10^{14}\ h^{-1} M_\odot$ and $\approx
1.0 \ h^{-1}$ Mpc, respectively).  In this case the kpSZ signal comes
only from the most internal regions (one third of virial radius) with
peaks up to few tens of $\mu$K.  Again the maps for the two
non-radiative simulations are very similar, with no significant
difference between the {\it ovisc} and {\it lvisc} models.  For the
simulations including cooling, star formation and feedback the typical
values of $\Delta T/T$ at the centre are smaller. The reduction is
particularly evident for the {\it csfc} model: unlike for the cluster
g51, in the case of g6212, given its smaller mass, the thermal
conduction is more efficient in smoothing down the signal.  Comparing
the results shown in Figs.~\ref{fig:g51_phys} and
\ref{fig:g6212_phys}, we find that, as expected, the different ICM
modeling has larger effects on smaller clusters. However, this
tendency is partially counterbalanced by the fact that the kpSZ signal
is mostly coming from the cluster core and low-mass clusters are more
concentrated than high-mass systems.

\begin{figure*}[!t]
  \centering
  \includegraphics[width=7cm]{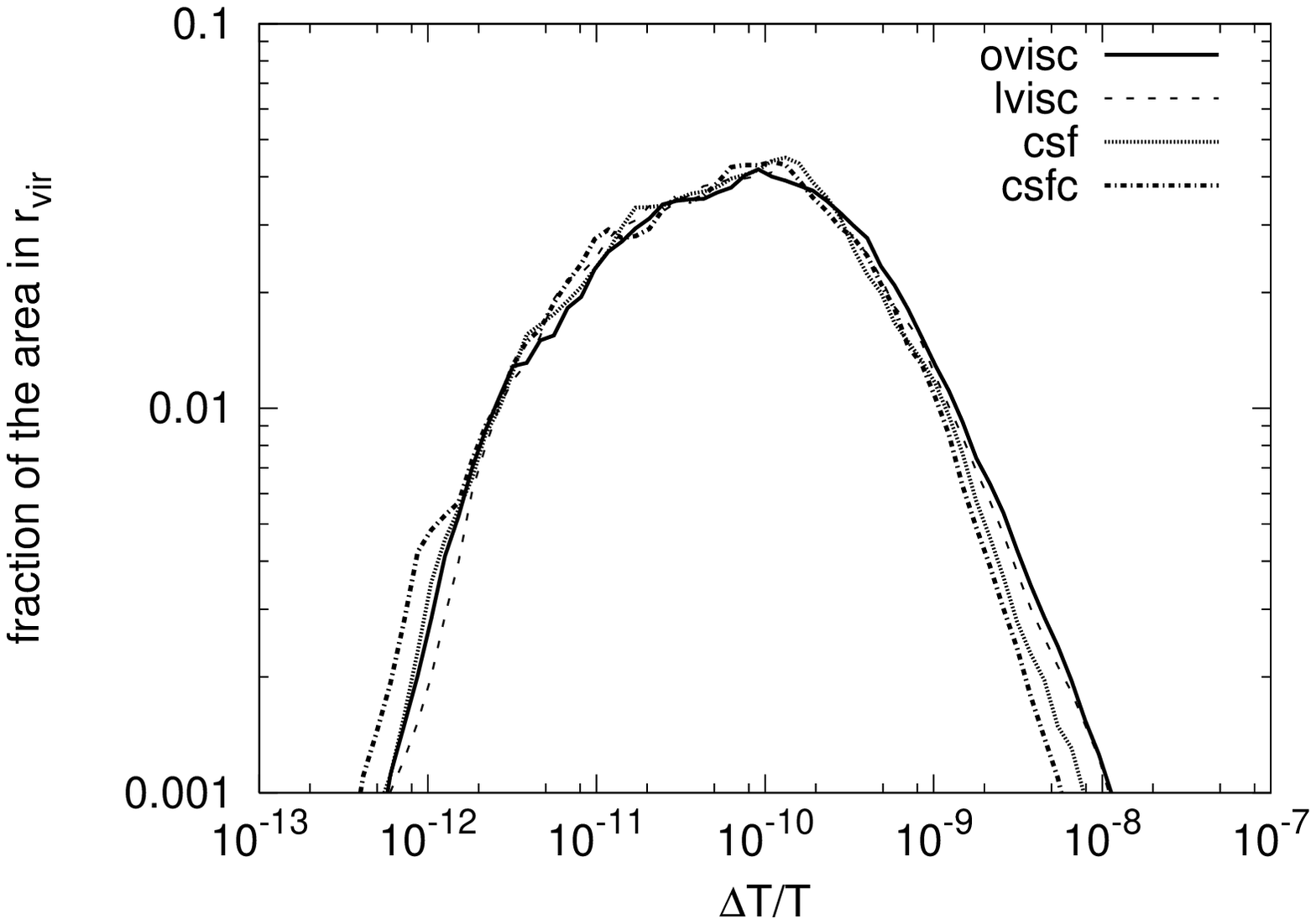}
  \includegraphics[width=7cm]{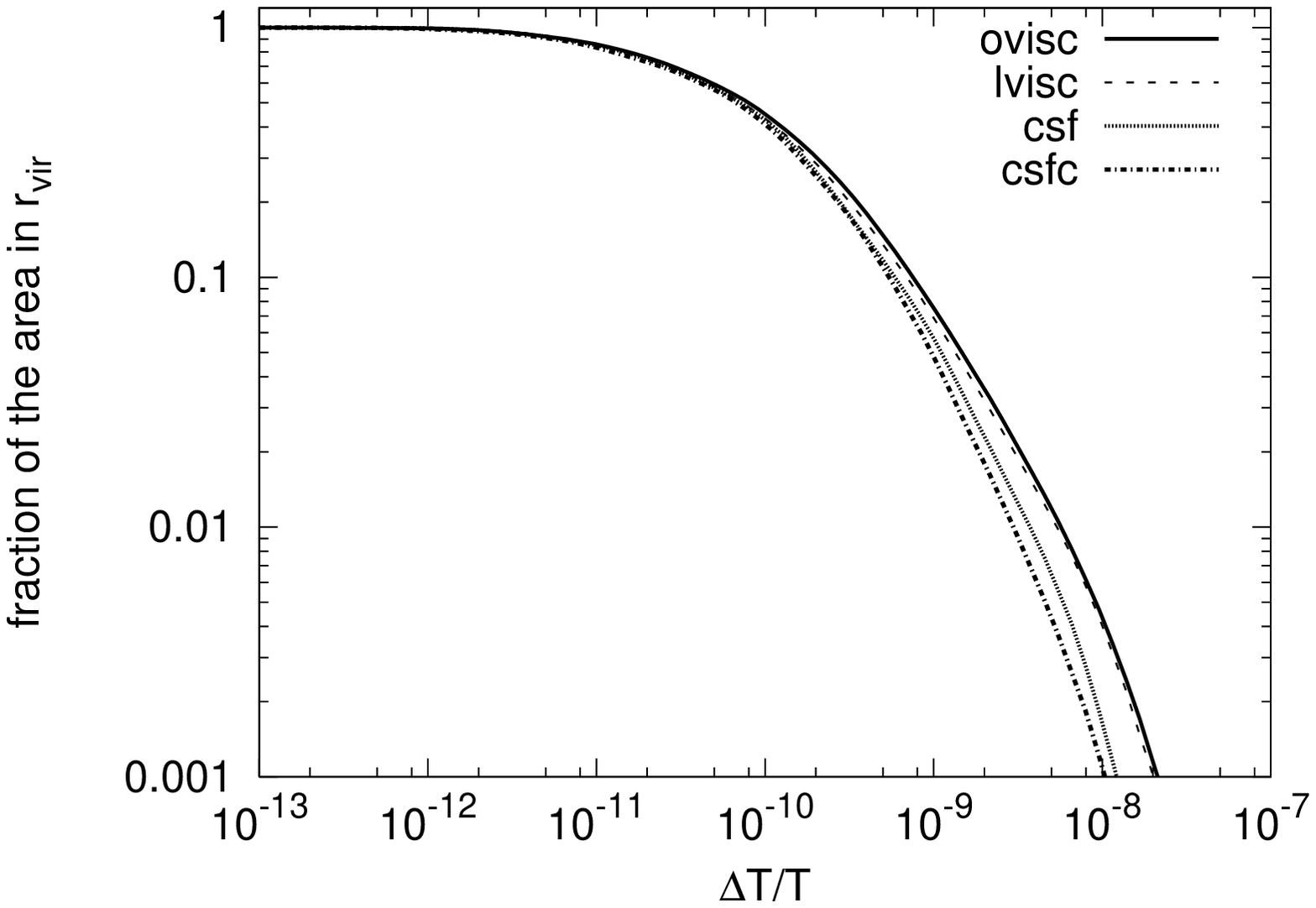}
  \caption{The differential (left panel) and cumulative (right panel)
  distributions of the polarization signal, as measured using all
  projections of the 9 simulated galaxy clusters belonging to our
  second sample. The different lines refer to results for simulations
  with different ICM modeling, as indicated in the panels.  } 
\label{fig:cum_statistica_fis}
\end{figure*}

A more quantitative picture can be drawn considering the statistical
properties of total sample.  In Figure \ref{fig:cum_statistica_fis} we
show the distributions of polarization amplitude for each of the four
ICM modeling here considered. In particular the curves represent the
median of the 27 maps and refer to the differential and cumulative
distributions (left and right panels, respectively).  As previously
pointed out, the differences between the {\it ovisc} and {\it lvisc}
models are negligible. The radiative processes acting in the {\it csf}
model produce some changes mostly in the high-signal tail, where the
polarization amplitude is reduced by at most a factor of 2.  A further
small signal reduction is given by the switching on of the thermal
conduction in the {\it cfsc} model. However, these differences are
involving only few per cent of the cluster total area.

\section{Conclusions}\label{sec:pol_conclusion}

We studied the possibility of measuring the dynamics of the ICM during
galaxy cluster mergers using the kinematic Sunyaev-Zel'dovich
polarization (kpSZ) induced on the CMB \citep{sunyaev1980,
sazonov1999}. This effect takes origin from the single scattering of
moving electrons: its dependence on the square of the tangential
electron velocity makes it a suitable probe to investigate the
kinematic properties of forming cosmic structures.

We used the outputs of 53 high-resolution hydrodynamical N-body
simulations to create realistic maps for this observable, which have
been statistically analysed, focusing the attention on the
polarization signal induced by in-falling blobs.  We demonstrated that
the amplitude of the cluster signal can vary up to one order of
magnitude depending on the internal cluster dynamics.

We found that the median kpSZ signal of the whole population is about
$10^{-8}\,K$, which could be observed by the upcoming millimetric
observatories.  In addition we showed that major merger events in
unrelaxed clusters can produce a polarization amplification of one
order of magnitude, with values up to $10^{-7}\,K$.  These high peaks
in the kpSZ signal are however related to fast in-falling blobs,
corresponding to small angular scales: consequently, an high
instrumental resolution (i.e. small aperture beam) is fundamental to
avoid to smear out the signal.

Our results show that the merging activity is of fundamental
importance to estimate the three-dimensional velocity power spectrum
of non-linear structure through the observations of the kpSZ and SZ
signals.  In fact, the amplification given by internal motions could
be dominant with respect to the signal originated by the bulk velocity
of clusters, resulting in a non-negligible source of noise for such
measurements.  This is also true for the estimation of the CMB
quadrupole from the observation of the polarization induced by the
intrinsic CMB quadrupole $P_Q$.  This makes necessary both a robust
component separation based on multi-frequency observations and a
careful selection of the clusters to be observed, which should be as
relaxed as possible.

\acknowledgements{
Computations have been performed by using the IBM-SP4/5 at CINECA
(Consorzio Interuniversitario del Nord-Est per il Calcolo Automatico),
Bologna, with CPU time assigned under an INAF-CINECA grant.  We
acknowledge financial contribution from contract ASI-INAF I/023/05/0
and INFN PD51. We thank Massimo Meneghetti and Daniele Dallacasa for
useful discussions. }

\end{document}